\documentclass[aps,prb,showpacs,byrevtex,showpacs,amsmath,amssymb,twocolumn]{revtex4}
\usepackage{epsfig}
\usepackage{graphicx}
\begin{document}
\title{From an insulating to a superfluid pair-bond liquid}
\author{M. Cuoco$^1$ and J. Ranninger$^2$}
\affiliation{$^1$ Laboratorio Regionale SuperMat, INFM-CNR, Baronissi (Salerno), Italy \\
Dipartimento di Fisica ``E. R. Caianiello'', Universit\`a di
Salerno, I-84081 Baronissi (Salerno), Italy\\
$^2$Centre de Recherches sur les Tr\`es Basses Temp\'eratures
 associ\'e \`a l'Universit\'e Joseph Fourier,
 C.N.R.S., BP 166, 38042 Grenoble-C\'edex 9, France}
\begin{abstract}
We study an exchange coupled system of itinerant electrons and
localized fermion pairs resulting in a resonant pairing formation.
This system inherently contains resonating fermion pairs on bonds
which lead to a superconducting phase provided that long range
phase coherence between their constituents can be established. The
prerequisite is that the resonating fermion pairs can become
itinerant. This is rendered possible through the emergence of two
kinds of bond-fermions: individual and composite fermions made of
one individual electron attached to a bound pair on a bond. If the
strength of the exchange coupling exceeds a certain value, the
superconducting ground state undergoes a quantum phase transition
into an insulating pair-bond liquid state. The gap of the
superfluid phase thereby goes over continuously into a charge gap
of the insulator. The change-over from the superconducting to the
insulating phase is accompanied by a corresponding qualitative
modification of the dispersion of the two kinds of fermionic
excitations. Using a bond operator formalism, we derive the phase
diagram of such a scenario together with the elementary
excitations characterizing the various phases as a function of the
exchange coupling and the temperature.
\end{abstract}
\pacs{74.20.-z, 74.20.Mn, 03.75.-b}  \maketitle
\section{Introduction}
The evolution of pairing correlations and, the related to it,
onset of phase coherence in low dimensional systems is at the
center of intense theoretical investigations.\cite{phil03} This
activity concerns systems such as: (i) high temperature cuprate
superconductors with their spin/charge pseudogap
phenomenon,\cite{lee06} (ii) low temperature superconducting
materials which can be driven towards insulating or metallic
phases via some extrinsic/intrinsic
mechanisms,\cite{Jae89,Eph96,Mas01} (iii) ultra-cold gases of
fermionic atoms in the cross-over regime between a BCS state and a
Bose-Einstein condensate, controlled by a Feshbach
resonance\cite{Joc03}.

In this paper, we shall investigate systems where bosonic resonant
pairs form in an ensemble of itinerant uncorrelated fermions. The
dynamics of such a boson-fermion exchange coupled system is
characterized by two competing processes: A) a local exchange
between a localized bound pair of fermions  and a pair of
itinerant uncorrelated fermions, B) a non-local single particle
hopping of the itinerant fermions between nearest neighbor sites.

The exchange between the localized bound pairs (acting as hardcore
bosons) and the fermionic itinerant particles creates a local
quantum superposition with bonding and antibonding resonant pair
configurations given by $|2,i\rangle_{\pm} = \frac{1}{\sqrt
2}(\rho^+_i \pm \tau^+_i)|0\rangle$. $\rho^+_i$ and $\tau^+_i =
c^{\dagger}_{i \uparrow}c^{\dagger}_{i \downarrow}$ denote
respectively the creation operators of the two constituents of
those local bound pair states, i.e. for  hardcore bosons and pairs
of itinerant fermions on site $i$. In the atomic limit, these
states are separated by an energy difference equal to twice the
boson-fermion exchange coupling, with the bonding state being
energetically favorable.

The local quantum structure of the bonding configuration, responsible
for inducing local resonant pairing among the fermions, has at the same
time a hindering effect in establishing a long range  superconducting
spatial correlations of such resonating pairs.
\begin{figure}[tbp]
\begin{center}
\includegraphics[width=0.3\textwidth]{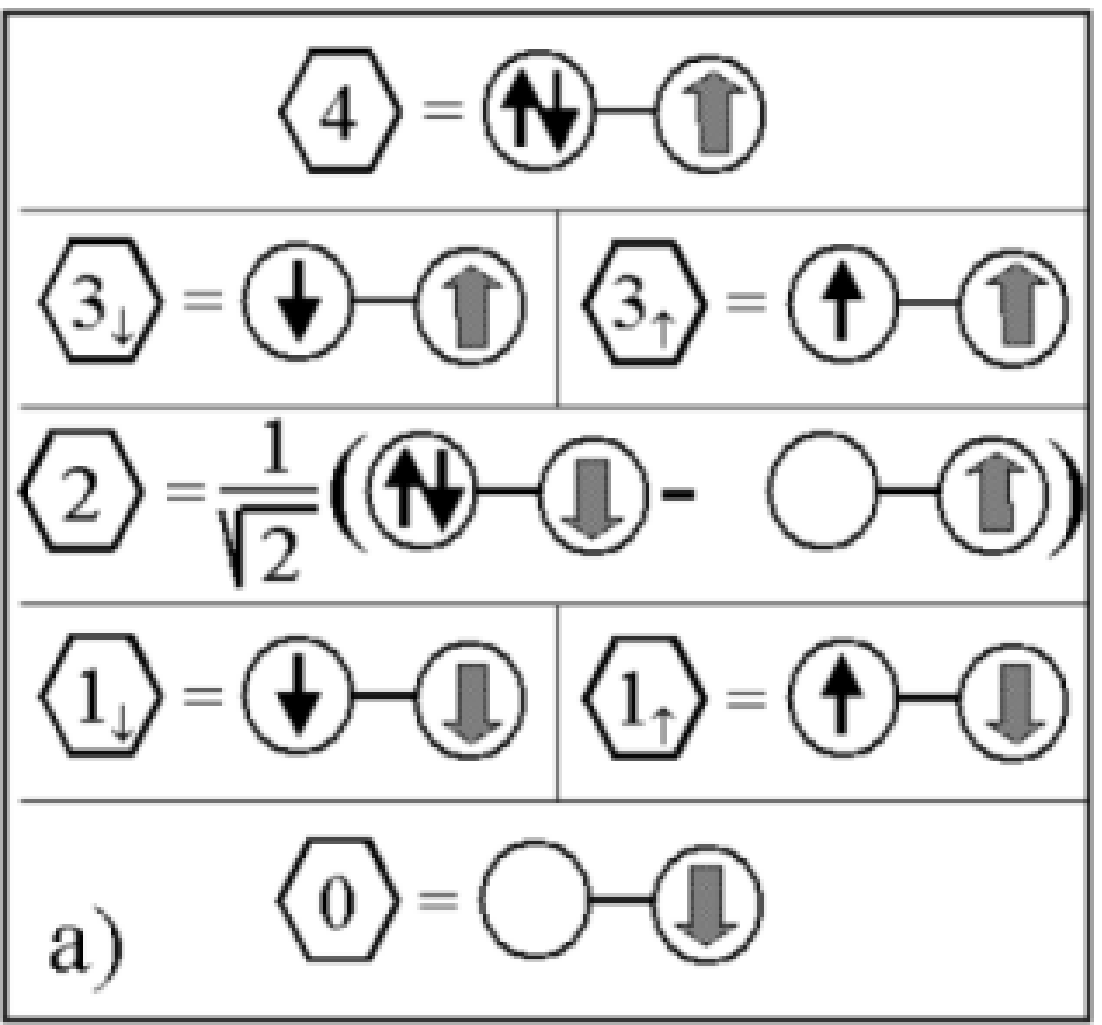}
\includegraphics[width=0.3\textwidth]{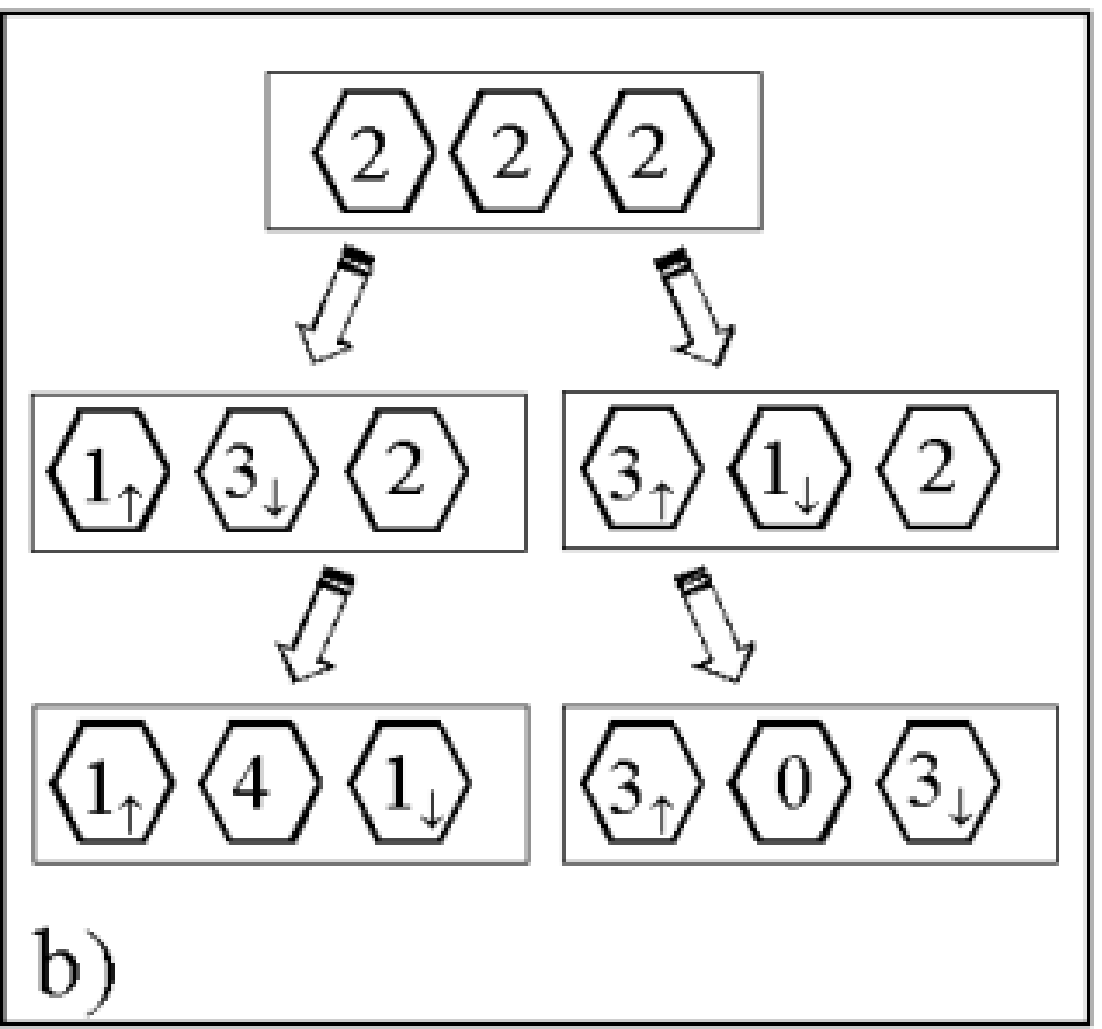}
\end{center}
\caption{The a) panel provides the description of the possible
configurations on the single bond (hexagon). The bold  arrow is
associated with the pseudospin for the original localized bound
pairs. The b) panel indicates the possible mechanisms of
dissociation starting from a representative string of three
neighbor bonding bosons. The successive two steps for that are: i)
the formation of two kinds of pairs, $1$ involving single
individual fermions on a bond and the other made up with charges
$3$ due to individual fermions attached with a bound pair on a
bond and subsequently ii) the bond fermions pairing up and
resonating with the bond-holes and the double-pairs-bonds.}
\label{sketch}
\end{figure}
Spatial correlations between the bonding/antibonding pairs are
built up via other bosonic and fermionic bond configurations which
form the local Hilbert space which has a very rich structure (see
Fig. 1a). There are two fermion-like configurations which result
from attaching to the vacuum state one individual fermion
(configuration $''1_{\sigma}''$), respectively such an individual
fermion together with a localized bosonic bound fermion pair
(configuration $''3_{\sigma}''$). Moreover, in this bond Hilbert
space we have completely empty bonds - bond-holes $''0''$ - and
completely filled ones - double-pair-bond states $''4''$. The
fermionic objects are itinerant. Their mobility is due to the
creation/annihilation of bonding (respectively anti-bonding) and
bond-holes (respectively double-paired bonds).

Our main aim in this study is to show that, due to the internal
degrees of freedom of the two components (localized hardcore
bosons and itinerant fermions), there are two distinct  energy
scales in this problem: one controlled by the dynamics of the
bonding pairs and one by the dynamics of the bond-holes and
double-paired bonds (see Fig. 1b). Depending on the ratio between
the exchange coupling strength and the hopping amplitude, the
dynamics of the two kinds of fermionic excitations  can lead to
either i) their pairing up in a superconducting state where
simultaneously a coherent state with bonding pairs, bond-holes and
double-paired bonds occurs, or ii) an insulating state with
coherent bonding pairs and with zero amplitude of the
bond-holes/double-paired bonds (see Fig.1b).

The presence of bonding pairs does not guarantee by itself the
occurrence of long range superconducting correlations between
their paired constituents. They have to dissociate in order to
induce a superconducting state. This process of dissociation
involves two possible channels for the dynamics of this resonant
pairing scenario (see Fig.\ref{sketch}b): i) the single fermions
and the composite ones (constituted of a fermion-boson pair)
hybridize and their itinerancy sustains a coherent liquid of
bonding pairs, ii) the two kinds of fermions pair up and  resonate
with the bond-holes and double-paired bonds and via that result in
a superconducting state.

Within this context, our target is the following: I) determine the
evolution of the superconducting state into an insulating state,
both quantum and thermally driven, II) analyze the nature of the
excitation spectrum as a function of the exchange coupling.

The outline of the paper is the following: in Sec II we present
the scenario for resonant pairing, on the basis of three examples
actively studied in the literature. In Sec. III we sketch the bond
operator formalism and adapt it to the boson-fermion problem.
Sec. IV is devoted to the derivation of the phase diagram as a
function of exchange coupling and temperature features. In Sec.
V, we present the evolution of the excitation spectrum as one
tunes the ground state from a superconductor to an insulator. The
last section VI is reserved for the summary, conclusions and
outlook.
\section{The Boson-Fermion scenario}
There is a variety of different physical systems where fairly
localized bound states are quasi degenerate with itinerant states
of their constituents. They can be paraphrased in terms of a
two-component system composed of localized bosonic pair states,
itinerant fermionic quasi-particles and a local exchange
interaction between the two. Such a scenario is described by the
so called boson-fermion model (BFM) Hamiltonian
\begin{eqnarray}
&H& =(D-\mu )\sum _{i\sigma }c_{i\sigma }^{\dagger}
c_{i\sigma }^{\phantom \dagger}
+(\Delta _{B}-2\mu )\sum _{i}(\rho_{i}^{z}+\frac{1}{2})\nonumber\\
&-&\sum _{i\neq j,\:\sigma } t_{ij} (c_{i\sigma }^{\dagger}
c_{j\sigma }^{\phantom \dagger} +H.c.)+ g\sum _{i}\left(
\rho^{+}_{i}\tau^{-}_{i} + \rho^{-}_{i}\tau^{+}_{i} \right),
 \label{Ham}
\end{eqnarray}
where $g$ is the strength of the exchange interaction and $t$ the
hopping integral for the itinerant fermions, which is assumed here
to be different from zero only for nearest neighbor sites. The
band half-width (which will serve as energy unit) is $D = zt$, $z$
being the coordination number of the underlying lattice. The
energy of the bound fermion pairs is denoted by $\Delta_B$. The
number  of the ensemble of bosons and fermions being conserved,
$n_{tot}=n_{F\uparrow} + n_{F\downarrow}+2 n_B$, implies a common
chemical potential $\mu$ for both subsystems. $n_B$ and
$n_{F\sigma}$ are the occupation numbers per site of the hard
core-bosons and of the fermions with spin $\sigma =
\uparrow,\downarrow$. In the present study, we restrict ourselves
to the analysis for the fully symmetric half-filled band case (
$\Delta_B \equiv 0$), which  in such a two-component system means
$n_{tot}=2$. The annihilation (creation) operators for the
fermions with a spin $\sigma$ at a certain site $i$ are given by
$c^{(\dagger)}_{i \sigma}$, those for the hard-core bosonic
fermionic bound pairs by  the pseudospin-$\frac{1}{2}$
$\rho^{\pm}_i$ and similarly those for the pairs of itinerant
uncorrelated fermions by $\tau^{+}_i=c_{i \uparrow}^{\dagger} c_{i
\downarrow}^{\dagger}$, $\tau^- = c_{i \downarrow} c_{i
\uparrow}$.

This model was introduced originally by one of us (JR) in an
attempt to capture the salient features of polaronic systems in
the intermediate coupling regime between adiabaticity and
anti-adiabaticity, but has turned out subsequently to be of much
more general relevance and applicability. Among others, such a scenario
has led to the prediction of the charge pseudogap features
in the high $T_c$ cuprate superconductors\cite{Ranninger-95},
without having to invoke any particular microscopic mechanism
for that. We shall discuss briefly three representative examples
where such a scenario seems to be relevant.

(i) In a system with strong local electron-lattice coupling we
have the formation of small polaronic charge carriers, which in
general will exist in form of localized bipolarons. If their
binding energy exceeds the bandwidth of the itinerant electrons in
an undeformed lattice, this can give rise to bipolaronic
superconductivity - an extremely fragile state of a phase
fluctuation controlled condensate of bosonic tightly bound
electron pairs -  whose existence in real materials has yet to be
experimentally verified. However, if the binding energy of such
localized bipolarons is such that it overlaps with the continuum
of the bare electron band states, this will result in an exchange
between such bound pairs and pairs of uncorrelated itinerant
electrons close to the chemical potential. The microscopic
mechanism for that exchange arises from large quantum fluctuations
of the lattice displacements in the immediate vicinity of the
charge carriers. Such local lattice displacements fluctuate
between essentially undeformed and much deformed lattice
environments which, as a result, periodically capture and release
electron pairs on small local cluster (acting as effective sites
on a lattice) made up of atoms and their associated ligand
environements\cite{Ranninger-Romano-06}.

(ii) In a system with strong local electronic correlations, as
described by the Hubbard model close to the half-filled band case,
hole pairing occurs on small plaquette
clusters.\cite{White-Scalapino-97} This, coexisting with triplet
spin pairs, destroys the antiferromagnetic long range order,
giving rise to a spin liquid state made out of singlet hole pairs,
which conceivably can condense into a superfluid
state.\cite{Anderson-87} One of the possible mechanism for that is
the hopping of the singlet hole pairs between neighboring
clusters. A more efficient way to get this condition is by
exchanging the local bound hole pairs on a plaquette with single
holes on neighboring plaquettes. This mechanism results in an
effective exchange coupling between bosonic bound hole pairs and
pairs of uncorrelated fermionic holes on adjacent plaquettes,
which at the end assure the itinerancy in the system. Such a
scenario has been analyzed on the basis of a kind of real space
renormalization group technique, where the initial Hubbard model
can be rephrased in terms of an effective BFM, albeit with
additional terms carrying the information of the underlying
antiferromagnetic short range
interaction.\cite{Altman-Auerbach-02}

(iii) In a system with an optically trapped gas of ultracold
fermionic atoms (studied in connection with the cross-over between
a BCS state of weakly interacting fermionic atoms and a
Bose-Einstein condensation of tightly bound states of them) one
can monitor the strength of the interaction between those fermions
- sweeping it from a repulsive to an attractive interaction - via
a so called Feshbach resonance
mechanism\cite{Feshbach-58,Timmermans-99} - under the effect of a
magnetic field. This mechanism is based on binary collision
processes in which the inter-atomic interaction depends on the
electronic spin configuration of the pair. For an overall
electronic spin triplet state (avoiding the Coulomb repulsion)
this leads to a weakly bound state, while a singlet configuration
leads to scattering states. Two incident atoms in a singlet
configuration can enter into a resonance with such a weakly bound
triplet configuration molecular state when their respective
energies are close to each other. This is achieved by flipping the
electron spin on one of the two atoms via a hyperfine interaction,
thus acquiring the necessary triplet configuration to bind them
momentarily into a pair. By applying a magnetic field during such
binary scattering processes one can change the relative position
of the energy levels of the two electron spin configuration and
thus enter in a resonance regime where they are quasi degenerate.
Such a situation can be then described by a phenomenological model
such as the BFM.


\section{Bond operator representation}
To study the scenario for resonant pairing, sketched above, by
explicitly taking into account the interplay between the bosonic
bonding pairs and the processes linking them to the single
particle fermionic entities, we make use of the bond operator
formalism and adapt it to the present BFM system. This approach
turns out to be particularly appropriate in treating situations
where there is a natural pairing in form of dimers in the ground
state, which is  either imposed by the Hamiltonian (as in our
case) or by a spontaneous symmetry breaking. The bond operator
theory has been successfully designed and applied in different
contexts such as for antiferromagnets\cite{Sach90,Chub91},
spin-ladder\cite{Gopa94}, doped antiferromagnets\cite{Park01},
bilayer quantum Hall\cite{Deml99}, and Kondo lattice
systems\cite{Jure01}.

Introducing the  bond operator formalism for the BFM, we start
from a bond on each lattice site being made up of fermionic and
bosonic configurations and then express the original fermionic and
bosonic operators $c^{(\dagger)}_{i \sigma}, \rho^{\pm}_i,
\tau^{\pm}_i$ in terms of the new  basis. Concerning a single
local bond, one has eight possible configurations to start with.
Similar to the case of a spin insulator, where one is introducing
the singlet and triplet boson operator, here we have four boson
operators that describe two-particle configurations in form of
{\it bonding-bonds} and {\it antibonding-bonds}  and zero- as well
as four-particle configurations in form of {\it hole-bonds} and
{\it double-paired-bonds}.  The two-particle bonding and
antibonding configurations can be expressed in terms of
pseudospin-$\frac{1}{2}$ operators $\tau^{+}_i,\rho^+_i$ for the
uncorrelated fermions and the tightly bound ones. The  zero- and
four-particle configurations are described by $|0\rangle$ and
$\rho^+_i\tau^+_i |0\rangle$ respectively. We thus have to
consider four bond boson operators $b^{(\dagger)}$,
$a^{(\dagger)}$, $f^{(\dagger)}$, and $d^{(\dagger)}$ (see Fig.
\ref{sketch}a) with
\begin{eqnarray}
b^{\dagger }|v\rangle &=&\frac{1}{\sqrt{2}}\left[
\rho ^{+}-\tau ^{+}\right] |0\rangle  \nonumber\\
a^{\dagger }|v\rangle &=&\frac{1}{\sqrt{2}}\left[ \rho ^{+}
+\tau ^{+}\right] |0\rangle \nonumber \\
f^{\dagger }|v\rangle  &=&\rho ^{+}\tau ^{+}|0\rangle \nonumber \\
d^{\dagger }|v\rangle  &=&|0\rangle.
\end{eqnarray}
$|0\rangle=|0_f\rangle \otimes |0_b\rangle$ denotes the boson-fermion
vacuum, i.e., the state with no fermions ($c_{\sigma}|0\rangle = 0$) and no
bosons present ($\rho^-|0\rangle =0 $).
$|v \rangle$ designates a corresponding new vacuum
state in which neither bond-bosons nor bond-fermions are present. The
remaining configurations are four bond-fermion operators describing
individual fermionic states and states where such fermions are attached
to a boson on a particular bond:
\begin{eqnarray}
h_{\sigma }^{\dagger }|v\rangle &=&c_{\sigma }^{\dagger }|0\rangle \nonumber \\
s_{\sigma }^{\dagger }|v\rangle &=&\rho
^{+}c_{\sigma }^{\dagger }|0\rangle  \,.
\end{eqnarray}
The operators $h^{(\dagger)} ,s^{(\dagger)}$ obey the canonical fermion
commutation relations, while
$d^{(\dagger)}, f^{(\dagger)}, b^{(\dagger)}, a^{(\dagger)}$ have boson
commutation rules. In this
representation, the total number of states in the Hilbert space of
these four bond bosons and four bond fermions is much larger than the eight
configurations allowed by the BFM. To limit the
kinematics to the physical subspace one has to impose the closure
relation on the bond given by the constraint:
\begin{eqnarray}
f^{\dagger}f+\sum_{\sigma}s_{\sigma }^{\dagger}s_{\sigma
}+b^{\dagger}b+a^{\dagger}a + \sum_{\sigma}h_{\sigma}^{\dagger}
h_{\sigma}+d^{\dagger
}d=\mathrm{1}\,. \label{costrain}
\end{eqnarray}
In the subspace limited by the relation (Eq. \ref{costrain}), one then
derives the expressions for the original fermion and
hard-core bosonic operators in terms of bond operators.

Let us first consider the fermion creation operator with spin
$\sigma$. Making use of the closure relation (Eq. \ref{costrain})
and considering all possible transitions induced by the single
particle fermionic operator between the different bond quantum
configurations, we obtain the following expression:
\begin{equation}
c_{\sigma }^{\dagger }=p_{\sigma }f^{\dagger }s_{-\sigma }+\frac{1}{\sqrt{2}}%
s_{\sigma }^{\dagger }\left( b+a\right) +\frac{p_{\sigma
}}{\sqrt{2}}\left( a^{\dagger }-b^{\dagger }\right) h_{-\sigma
}+h_{\sigma }^{\dagger }d \,,
\label{c}\end{equation}
where $p_{\sigma }=(+1,-1)$ for $\sigma =(\uparrow,\downarrow)$. In a similar
way one obtains for the creation operator of
a hard core boson:
\begin{eqnarray}
\rho ^{+}=\frac{1}{\sqrt{2}}f^{\dagger }(a-b)+\sum_{\sigma
}s_{\sigma }^{\dagger }h_{\sigma }+\frac{1}{\sqrt{2}}(b^{\dagger
}+a^{\dagger })d \,.
\end{eqnarray}
This, subsequently leads to the following expressions for the density
operators for the bosons and fermions:
\begin{eqnarray}
\rho ^{+}\rho^- &=& \frac{1}{2}(a^{\dagger }a+b^{\dagger
}b+a^{\dagger }b+b^{\dagger }a)+f^{\dagger }f+\sum_{\sigma
}s_{\sigma }^{\dagger }s_{\sigma } \nonumber\\
c_{\sigma }^{\dagger }c_{\sigma } &=& \frac{1}{2}(a^{\dagger
}a+b^{\dagger }b-a^{\dagger }b-b^{\dagger }a)+f^{\dagger}f+\left(
s_{\sigma }^{\dagger }s_{\sigma }+h_{\sigma }^{\dagger }h_{\sigma
}\right). \,\nonumber \\
\end{eqnarray}
Finally, the expression for the pair exchange on the bond between
the initial fermions and bosons reduces in the new representation
to just the difference between the number of bonding and
antibonding bosons, that is:

\begin{eqnarray}
\rho ^{+} \tau^-  + \tau^{+}\rho^-=a^{\dagger
}a-b^{\dagger } b \,.
\end{eqnarray}
\section{Mean-field phase diagram}
Let us now apply the bond operator method to study the spectral
properties of the BFM, in conjunction with its various possible
phases. We proceed by looking for solutions where the bosons associated
with the bonding-bonds
$b^{(\dagger)}$, antibonding-bonds $a^{(\dagger)}$,
the bond holes $d^{(\dagger)}$, and double-paired-bonds
$f^{(\dagger)}$ are condensed.
This means that in the Hamiltonian, Eq.\ref{Ham}, after having
transformed the original fermion and boson operators into the new
bond operators, one is
decoupling the quartic terms in a way such that the expectation value
of the linear and bilinear bond boson operators have a non-zero
amplitude, $\langle o \rangle = \bar o$, with $\bar o= \bar b, \bar a,
\bar f, \bar d$. The coherent state  which then emerges in the most
general case has the following local structure:
\begin{eqnarray}
|\psi_c \rangle=(u_v+u_b b^\dagger+u_a a^\dagger+u_d d^\dagger+u_f
f^\dagger)|v\rangle,
\end{eqnarray}
\noindent where the finite amplitudes of the various coefficients $u$ are
related to the condensate of the corresponding bosonic degrees of
freedom. Let us, at this point,  make some general observations related to the
structure of the Hamiltonian in such a bond operator representation.
The $b$-bond bosons are always condensed at zero temperature
($\bar b \neq 0$), due to the local exchange and  the
hopping induced processes related to the creation of $d$-bond bosons
and $s$, $h$ bond fermions. Moreover,  the $d$-bond bosons
condense only if simultaneously the $f$-bond bosons condense and in our
present analysis concentrating on the particle-hole symmetric half-filled
band case, we have
($\bar d = \bar f \neq 0$). In the present mean field approach
the $a$-bond bosons have gapped excitations due to the exchange
splitting with respect to the $b$-bond bosons and hence
do not contribute to a coherent state in the lowest energy configuration.

Let us now examine  the possible solutions at zero
temperature as a function of the exchange coupling $g$. Our main
objective is to show how, above a critical value of the ratio
$g/D$, the system passes from a superconducting state with
simultaneous coherence of $b$-bond bosons and the $d$-bond as
well as $f$-bond bosons, to an insulating state where
only the $b$-bond bosons are condensed.

Before going into the details about such a calculation and the
results, let us point out certain aspects of the emerging bond
boson-fermion dynamics which can be envisaged on simple and quite
general physical grounds. As mentioned before, the $b$-bond bosons
are always in a coherent state at zero temperature, due the local
exchange and the quantum processes of double creation and
annihilation of nearest neighbor $b$-bond bosons. Hence, the U(1)
symmetry associated to the number conservation of the $b$-bond
bosons is generally broken in the entire regime of couplings $g$.
In the weak coupling regime, $b$-bond bosons have a large
dispersion which, together with coherent $d$-bond and $f$-bond
bosons, leads to a superconducting state. In the
intermediate/strong coupling regime, the $b$ bosons become more
populated, which is caused by an increased local binding energy.
This increase in the $b$,\,$d$ boson occupation leads to a
decrease of the $s$- and $h$-bond fermion densities, which
ultimately causes a reduction of the $b$ boson dispersion. Once
the $d$- and $f$-bond bosons are no longer condensed, the
coherence of the $b$-bond bosons is built up via processes
described by the terms $b_i b_j s^\dagger_{i\sigma}
h^\dagger_{j\bar{\sigma}}$($b^\dagger_i b^\dagger_j s_{i\sigma}
h_{j\bar{\sigma}}$) in the Hamiltonian.

Let us now consider the physics sketched above in terms of a mean-field
analysis of the BFM. After performing the corresponding decoupling procedure,
the Hamiltonian can be separated into three main parts:
\begin{eqnarray}
H=H_{t}+H_{ex}+H_0 \label{H1},
\end{eqnarray}
\noindent where $H_t$ is the hopping term, $H_{ex}$ the local pair
exchange, and $H_0$ a term including the contributions coming from bond bosons
in the chemical potential term and the constraint. Here, the
constraint is treated in an approximative way, where the corresponding Lagrange
multiplier is  taken as spatially homogeneous and  which has to be
determined variationally as a saddle point solution of the total free
energy. In our  mean-field approximation, we rewrite the first part of
Eq. \ref{H1} in a compact way by
introducing the vector operator $D_k=(s_{k,\uparrow},s_{-k,%
\downarrow}^{\dagger},h_{k,\uparrow},h_{-k,\downarrow}^{\dagger})$
and its hermitian conjugate $D_k^{\dagger}=(s_{k,\uparrow}^{\dagger},s_{-k,%
\downarrow},h_{k,\uparrow}^{\dagger},h_{-k,\downarrow})$ in terms of the
Fourier representation for the bond fermions. The single-particle excitations
are then determined by the following mean field Hamiltonian:
\begin{eqnarray}
H_t=\sum_{k} D_k^{\dagger} \, \hat{U}_k \,D_k+c,
\end{eqnarray}
where $c=2\lambda-4\mu$ is a constant which arises in the process of
ordering of the various bond fermionic operators. The 4x4 matrix $\hat{U}_k$
is given by:
\[
\hspace{-0.5cm}\left[
\begin{tabular}{ccccccc}
$\gamma^s_{k}$ & \thinspace  & $\Delta^s_k$ \thinspace  & $\gamma^{\bar{s}h}_{k}$ & \thinspace  & $\gamma^{\bar{s}\bar{h}}_{k}$ \\
&  &  &  &  &  &  \\
$\Delta^s_k$ & \thinspace  & $-\gamma^s_{k}$ \thinspace  &
$\gamma^{{s}h}_{k}$ & \thinspace  & $-\gamma^{\bar{s}h}_{k}$
\\
&  &  &  &  &  &  \\
$\gamma^{\bar{s}h}_{k}$ & \thinspace  & $\gamma^{s h}_{k}$
\thinspace  & $\gamma^{h}_{k}$ & \thinspace  & $\Delta^h_{k}$
\\
&  &  &  &  &  &  \\
$\gamma^{\bar{s}\bar{h}}_{k}$ & \thinspace  &
$-\gamma^{\bar{s}h}_{k}$ \thinspace  & $\Delta^h_{k}$ & \thinspace
& $-\gamma^{h}_{k}$
\end{tabular}
\right]
\]

\noindent $\epsilon_k=-t \sum_{\delta} \exp[i k\cdot {\delta}]$ is
the free particle spectrum of the original fermions, with $\delta$
designating the lattice vectors linking nearest neighbor sites.
$\gamma^{s}_{k}= \epsilon_{k} [0.5*(b+a)^2-f^2]-3 \mu +\lambda $,
$\gamma^{h}_{k}=\epsilon_k [d^2-0.5*(b-a)^2]-\mu +\lambda$ are the
single particle energy spectra for the $s$ and $h$ fermions.
$\gamma^{\bar{s}h}_{k}=\frac{\epsilon_k}{\sqrt 2}
\{[(b+a)d]-[\left( a-b \right) f]\}$ and $\gamma^{\bar{s}\bar{h}}=
\epsilon_k[(f\,d)\,+\frac{1}{2}\left( a^{2}-b^{2}\right) ]$ are
the particle/hole and particle/particle hybridization factors
between the $s$- and $h$-bond fermions. The pairing amplitudes for
the $s$- and $h$-bond fermions are
$\Delta^s_k=\epsilon_{k}\sqrt{2}[f(b+a)]$ and $\Delta^h_k=
\epsilon_{k}\sqrt{2}[d(a-b)]$, respectively.

Let us next investigate how the effective hopping of the $s, h$-bond
fermions depends on the amplitude and the density of the condensed
bosons. As one can see by inspection of the matrix $\hat{U}_k$,
the processes linked to the fermionic degrees of freedom are
strongly renormalized by the strength of the condensed
$b$-bond bosons. Concerning the dispersion of the $s, h$-bond fermions, one
can see that the contribution  of the density of the $d, f$-bond bosons
counteracts that of the $b$-bond bosons. The strength of the
pairing amplitude being proportional to the product of the $b$-bond
and $d, f$-bond boson condensate amplitudes, indicates the need of having
both types of bond bosons condensed in a state where the  $s, h$-bond fermions
are paired up. The hybridization between the $s$- and $h$-bond  fermions
manifests itself in both, the particle/particle and in the
particle/hole channel. While the amplitude $\gamma^{\bar{s}h}_{k}$
of the $s-h$ hybridization is linearly linked to the $b$-bond and
$d, f$-bond boson condensate, the processes related to the
particle/particle mixture depend on the effective density of the
$b$-bond bosons as well as the $d, f$-bond bosons. All those
processes compete with each other, resulting in either a
superconducting or an insulating state, depending on the value of $g/D$.

Let us next examine the exchange and the local contributions of the
BFM, which after a corresponding mean-field decoupling are given by:
\begin{eqnarray}
H_{ex}=&& L\, g (\bar a^2 - \bar b^2)\nonumber \\
H_{0}=&&-L\, \mu \lbrack 4\bar f^{2}+2(\bar b^{2}+\bar a^{2})]+\nonumber \\ && L\, \lambda
\lbrack\bar  f^{2}+\bar b^{2}+\bar a^{2}+\bar d^{2}-1]
\end{eqnarray}
and where $L$ indicates the total number of bonds. The procedure of the
present analysis is to look for a saddle point solution of the total free
energy  with respect to the bond boson amplitudes and the Lagrange
multiplier for the constraint.
The free energy is obtained after performing a Bogoliubov
rotation of the $s$, $h$-bond boson operators in the $H_t$ term, bringing
it into the diagonal form:
\begin{eqnarray}
H_t=\sum_{k}E_{\alpha \,,k}\,\eta _{\alpha ,k}^{\dagger }\,\,\eta
_{\alpha ,k}.
\end{eqnarray}
$E_{\alpha \,,k}$ ($\alpha =1,..,4$) are the eigen-energies of the
various Bogoliubov quasi-particles $\eta _{\alpha ,k}^{\dagger
}$($\eta _{\alpha ,k}$), obtained by diagonalizing the matrix
$\hat{U}_{k}$. They result from a unitary transformation of the
original fermions and thus have the same commutations relations as
$D_{k}^{\dagger}$($D_{k}$). The free energy is then given
by the following expression:
\begin{eqnarray}
F&=&-\frac{1}{\beta}\frac{1}{L}\sum_k \sum_{i=1,..4} \ln[2
\cosh(\frac{\beta}{2}
E_{i,k})]+E_0 \nonumber \\
E_0&=&g (\bar a^2-\bar b^2)-\mu \lbrack 4\bar f^{2}+2(\bar
b^{2}+\bar a^{2})]+
\nonumber \\
&&\lambda \lbrack \bar f^{2}+\bar b^{2}+\bar a^{2}+\bar d^{2}-1]
\,
\end{eqnarray}
\noindent with $\beta=1/(k_B T)$.

The saddle point solutions for the bond bosons together with the
Lagrange multiplier are obtained variationally by requiring the following
extremal conditions:
\begin{eqnarray}
&&\frac{\partial F}{\partial \lambda}=0 \nonumber \\
&&\frac{\partial F}{\partial \bar a} =\frac{\partial F}{\partial \bar b}=
\frac{\partial F}{\partial \bar d}=\frac{\partial F}{\partial \bar f}=0
\label{Eq.extrema}
\end{eqnarray}
together with the chemical potential being fixed via the relation
$n_{tot}=-\frac{\partial F}{\partial \mu }$.

\begin{figure}[tbp]
\begin{center}
\includegraphics[width=0.45\textwidth]{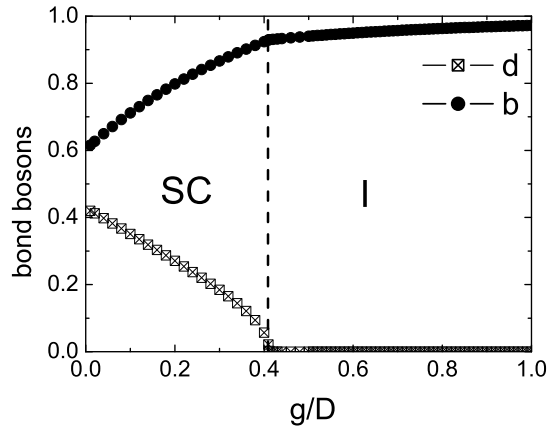}
\end{center}
\caption{Phase diagram at zero temperature as given by the
evolution of the bonding bosons and bond hole amplitude. SC and I
stands for a superconducting and insulating state respectively.}
\label{T0}
\end{figure}

In Fig.\ref{T0}, we report the evolution of the amplitude for the
condensed $b$-bond bosons at zero temperature as a function of
$g/D$. Throughout this paper we use a 2-dimensional fermionic
tight binding spectrum  $\epsilon_k=-(D/2)(\cos[k_x]+\cos[k_y])$
for the original fermions. The self-consistent solutions of the
non-linear Eqs. \ref{Eq.extrema} yield a zero amplitude for the
$a$-bond bosons, as anticipated above, and an equal amplitude for
the $d$- and $f$-bond bosons. The latter is a consequence of the
particle-hole symmetric case considered here. We also find a clear
separation between two distinct regions as the exchange coupling
is varied. Below a critical $g_{crit}/D \sim 0.4$ the system
allows a finite condensation of the $d, f$-bond  and the $b$-bond
bosons, which signifies a superconducting state. Yet, upon
increasing the exchange coupling, the amplitude of the $d, f$-bond
bosons gets reduced, until above $g_{crit}$, the ground state does
not contain  anymore any $d$-bond bosons. The transition is
continuous, merely showing a change of slope in the behavior of
$\bar b$.

A particular feature of this mean field analysis at $T=0$ is that
the order parameter $\bar d$, controlling the superconducting
phase, has a finite value as soon as the boson-fermion exchange
mechanism is switched on, i.e., for $g/D \neq 0$ however small it
might be. In the limiting case were the exchange coupling is zero,
the superconducting solution disappears because then $b$-bond and
$a$-bond bosons are equally populated. Such a degeneracy is
removed as soon as the exchange coupling is different from zero.
The $b$-bond bosons are now being able to condense while $a$-bond
bosons will not. Thus, in the present mean field treatment the
system finds the solution with a finite amplitude for the $b$-bond
and $d, f$-bond bosons as soon as the exchange is switched on at a
non-zero value. The size of the $b$-bond boson amplitude is then
related to the constraint and via that to the total filling under
consideration.

At this point we would like to make a few remarks on the weak
coupling regime. Since the mean-field analysis is based on a local
ansatz for the saddle point solution, we do expect that the
present approach is best suited for the strong-coupling limit. In
the weak coupling limit such a scheme of approximation is not
completely satisfactory. In that regime, the dynamics of the bond
bosons and their delocalized behavior have to be taken into
account in order to properly consider their phase and amplitude
fluctuations and the related to it consequences on the fermionic
subsystem. The bosonic fluctuations should renormalize down to
zero the value of the critical temperature in this small coupling
region. The finite amplitude of T$_c$ at $g/D \sim 0$, obtained in
our present mean-field analysis, indicates that the pairing
formation is driven by kinetics and an overestimation of the phase
locking with respect to the amplitude fluctuations of the bond
boson order parameter occurs. This physically unsatisfactory
feature is corrected by including the competition between the
phase and amplitude dynamics, as we have done in a recent work
where an effective phase action has been derived after integrating
out the fermionic degrees of freedom (see insert in Fig.
3).\cite{Cuo05} We do believe that, even in the weak coupling
regime, a BCS type of superconductivity should not occur since the
charge dynamics intrinsically involves the two types of $s-$ and
$h-$ bond fermions. A direct decoupling of the exchange
interaction in the starting representation of the BFM would hence
not be justified as the hard core bosons are localized fields and
thus in principle cannot have a phase coherent amplitude in the
transverse mode. In an attempt to include the mobility of the
bosons, in a self-consistent perturbative (weak-coupling limit)
approach, it has been shown that already above the superconducting
transition temperature one gets, apart from the regular free band
spectrum, two rather dispersion-less branches.\cite{Ran96} As we
will see below, such an excitation spectrum resembles that
obtained in the present approximation. Hence, within our saddle
point solution qualitatively correct features can be recovered
even in the weak-coupling regime.

Concerning the dynamics of the bosons, we should make a few
remarks about the possible nature and the character of the bosonic
excitations near the quantum phase transition between the
superconducting and the insulating state - in a frame that is
beyond our present approximative mean field scheme of
investigation. Focusing on the dynamics on the insulating side, we
observe that the fermionic excitations are gapped, while the
$b$-bond bosons may have low energy gapless excitations resulting
from their itinerancy. However, such excitations are not
$''charged''$, i.e.,  capable of creating a pair of fermions, or a
localized boson, since that would require at the same time the
presence of both $b$-bond as well as $d, f$-bond bosons. Hence,
two scenarios are possible as $g/D$ is increased: i) the $d,
f$-bond boson coherence and their density go to zero
simultaneously, ii) the phase coherence of the $d$-bond bosons
drops to zero before their density vanishes. Case i) corresponds
to a direct transition between the superconducting to an
insulating configuration with coherent $b$-bond bosons. Case ii)
allows to have an intermediate state, where a finite density of
$d, f$-bond bosons occurs in a state that is not condensed. In
this case, the current can be carried directly by the mobile
$b$-bond and $d, f$-bond bosons, and the presence of these low
energy excitations can induce a metallic-like state. This regime
is quite exotic, since the effect of the $a$-bond bosons in
activating itinerant pairs can induce a gap in the bosonic sector
and, in this way, result again in an insulating state. The picture
that finally could emerge is one of a system that undergoes first
a transition from a superconductor to a metallic-like phase (built
out of a state where incoherent itinerant $b$-bond  and $d,
f$-bond  bosons are present and which ultimately transits to an
insulating state where, (driven by the exchange coupling and due
to the constraint) there exist only condensed $b$-bond bosons and
no condensed $d, f$-bond bosons.

\begin{figure}[tbp]
\begin{center}
\includegraphics[width=0.45\textwidth]{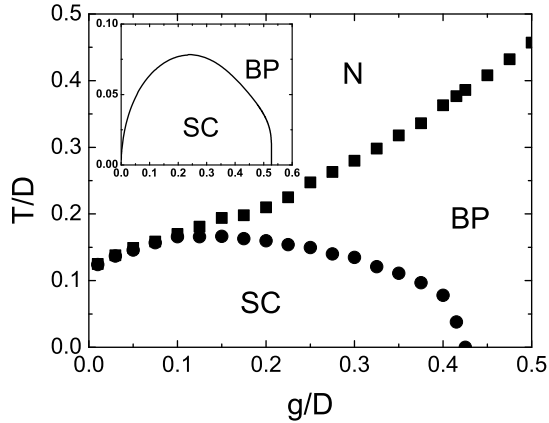}
\end{center}
\caption{Phase diagram at finite temperature as a function of
$g/D$ as given by the evolution of the $b$-bond boson and $d$-bond
boson amplitude. SC, BP and N stands for a superconducting, bond
coherent boson phase, and normal state, respectively. The insert
shows the evolution of the critical superconducting temperature as
obtained within a path-integral approach.\cite{Cuo05}}
\label{PDvsT}
\end{figure}

Let us now analyze  the behavior of the two coherent states as a
function of  temperature (see Fig. \ref{PDvsT}). In the structure
of the phase diagram one notices three regions that are
reminiscent of the configurations found at zero temperature. The
superconducting state (SC) is characterized by a non-zero
amplitude for the $b$- as well as $d, f$-bond bosons, the BP (bond
phase) by condensed $b$-bond bosons but a zero amplitude of the
$d, f$-bond bosons, while in the normal (N) high temperature phase
none of the bond bosons are condensed. As a consequence of the
competition between the $b$-bond  and $d, f$-bond boson
condensation, one has two regimes depending on the exchange
coupling in the region where SC is stable at zero temperature. For
values of the exchange coupling below $g^*/D=0.125$, the SC
critical temperature ($T_{bd}$) and that associated with the
$b$-bond boson condensation ($T_b$) are very close in amplitude so
that there occurs an almost  direct transition from the SC to the
N phase. In this limit, the scales of energy that controls the
$b$-bond  and $d, f$-bond dynamics are comparable. Going to a
larger exchange couplings ($g > g^*$), the characteristic
temperature marking the onset of SC is reduced due to the
increased population of the $b$-bond bosons. But contrary to that,
$T_b$ grows almost linearly with the coupling. Thus, outside the
quantum phase transition region there occurs a large region in
parameter space, where the system is not superconducting but is
still characterized by a coherent state of $b$-bond bosons.

\section{Evolution of the charge gap and of the spectral function from the superconducting to the insulating region}

In this section we analyze the excitation spectra both for the
$s$- and  $h$-bond fermions and for the original $c$ fermions. We
shall focus on the evolution of the spectral function when moving
from the superconducting to the insulating region in the phase
diagram. In the previous section, we have constructed the vector
$D_{k}$, with $D_{k,i}$ denoting its $i-th$ component. Due to its
structure, the time dependent correlation function for the vector
$D_{k}$ contains the information for the matrix Green's function
both for the diagonal and the off-diagonal anomalous part of the
$s$- and $h$-bond bosons. Using the relation  which connects the
$s$- and $h$-bond-fermions and the $b$-bond bosons with the
original $c$ type fermions, we are able to extract the spectral
function for the original fermions. For that, it is convenient to
introduce the following time dependent correlation function:
\begin{equation}
G_{k}^{ij}(\tau )=-\langle T\left( D_{k,i}(\tau )D_{k,j}^{\dagger
}(0)\right) \rangle
\end{equation}
where $i,j=1,..,4$ and $\tau $ is the Matsubara imaginary time,
$T$ the usual time ordering operator, and $\langle ...\rangle $
indicates the average at finite temperature. Due to the bilinear
structure of the Hamiltonian, we readily determine the Green
function via its equation of motion. We thus obtain the following
expression in matrix notation:
\begin{equation}
\hat{G}_{k}(\omega _{n})=(i\omega
_{n}\mathbf{1}-\hat{U}^{k})^{-1}\,.
\end{equation}
\begin{figure}[tbp]
\begin{center}
\includegraphics[width=0.45\textwidth]{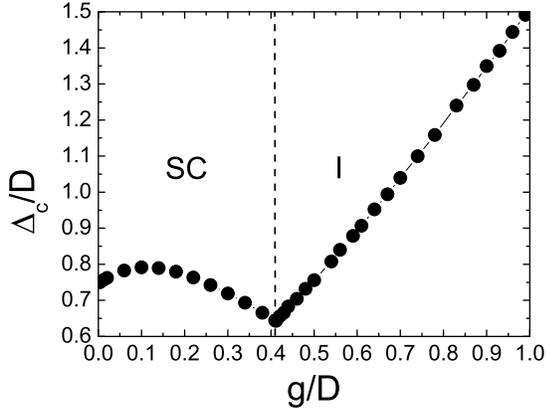}
\end{center}
\caption{Zero temperature evolution of the charge gap associated
with the excitation of a $c$ fermion as a function of the
exchange coupling.} \label{Gapvsg}
\end{figure}
The Green's function for the original $c$ operators, is consequently a
linear combination of the contributions of the different components
$G_{k}^{ij}(\tau )$  which are weighted by the various
amplitudes of the condensed $b$-bond bosons. This is because of  our mean
field approach where in the expression for the fermionic $c$ operators,
Eq. (\ref{c}), the various boson operators are replaced by their mean
field averages, i.e.,
\begin{eqnarray}
c_{i\sigma }^{\dagger } &=&p_{\sigma }\bar f s_{i\bar{\sigma}}+\frac{1}{%
\sqrt{2}}s_{i\sigma }^{\dagger }\left( \bar b+\bar a \right) +
\frac{p_{\sigma }}{\sqrt{2%
}}\left( \bar a - \bar b \right) h_{i\bar{\sigma}}+h_{i\sigma }^{\dagger }
\bar d \nonumber \\
c_{i\sigma } &=&p_{\sigma }\bar f s_{i\bar{\sigma}}^{\dagger }+\frac{1}{%
\sqrt{2}}s_{i\sigma }\left( \bar b + \bar a \right) +\frac{p_{\sigma
}}{\sqrt{2}}\left( \bar a - \bar b \right) h_{i\bar{\sigma}}^{\dagger
}+h_{i\sigma } \bar d \nonumber
\\
\end{eqnarray}

Expanding the expression for the time dependent Green's function
of the $c$ fermions $G_{k\sigma }^{c}(\omega _{n})$, we derive the
following structure for it:

\begin{eqnarray}
G_{k\sigma }^{c}(\omega _{n})= Tr [\hat{F}\cdot \hat{G}_{k}(\omega
_{n}) ]
\end{eqnarray}
where $Tr$ indicates the trace of the matrix indices, and
$\hat{F}$ is a 4x4 matrix whose coefficients depend on the
amplitude of the boson condensates.

\begin{figure}[tbp]
\begin{center}
\includegraphics[width=0.35\textwidth]{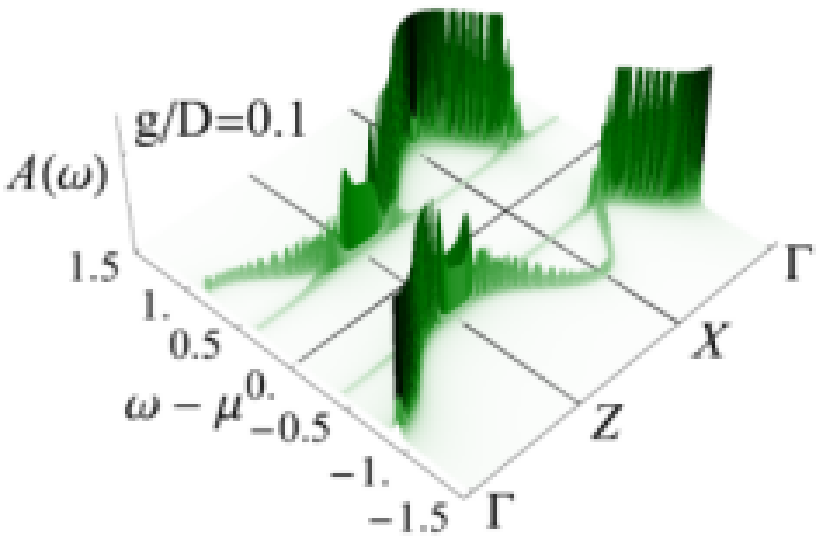}
\includegraphics[width=0.35\textwidth]{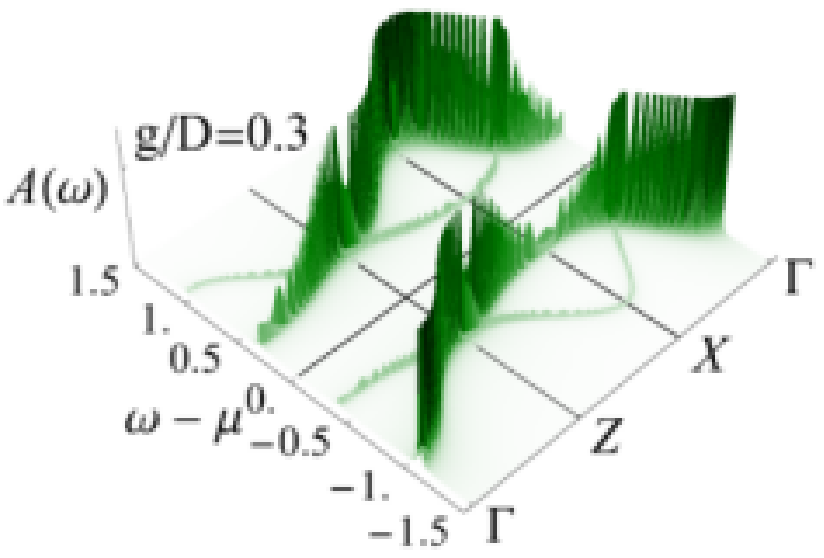}
\includegraphics[width=0.35\textwidth]{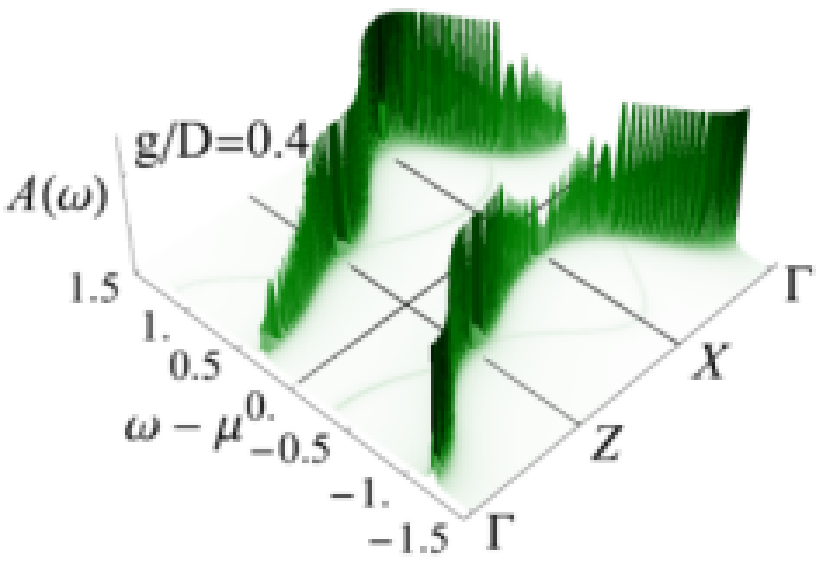}
\includegraphics[width=0.35\textwidth]{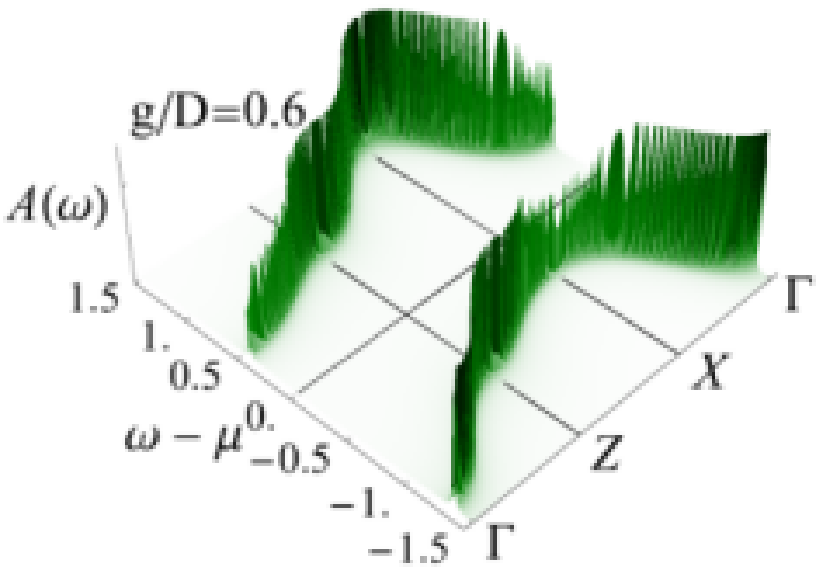}
\end{center}
\caption{(Color online) Dispersion of the Bogoliubov branches
$E_i$ and the related spectral weight at zero temperature for the
$c$-type excitation as a function of the exchange coupling. From
top to the bottom the exchange assumes the following values:
$g/D=0.1,0.3,0.4,0.6$, respectively. The points of the two
dimensional Brillouin zone are:
$\Gamma=[0,0],Z=[\pi,0],X=[\pi,\pi]$.} \label{SPvsg}
\end{figure}
In our investigation of the various phases of the BFM, let us
start with by examining the evolution of the charge gap, as
extracted from the density of states, $N_c(\omega)=\int d\epsilon
N(\epsilon) A^c_k(\omega)$, with $A^c_k(\omega)=(-1/\pi)
Im[G^{c}_k(\omega)]$ and $N(\epsilon)$ the bare density of states.
Both in the SC- and the I-phase the $b$-bond bosons are condensed.
Thus, the charge gap ($\Delta_c$) represents an estimate of the
energy required to break a $b$-bond boson and create a pair of
$h$- and $s$-bond fermions in case those bond fermions simply
hybridize or pair up. As one can see from Fig.\ref{Gapvsg},
$\Delta_c$ has a non-monotonous behavior as the exchange coupling
$g$ is varied from weak to strong coupling. For $g/D$ smaller than
about $0.15$, the energy to excite a single $c$-type fermion grows
slightly as a consequence of the concomitant effect of the
constraint, the hybridization between the $s$- and $h$-bond
fermions as well as the increasing pairing between them. Beyond
$g/D \simeq 0.15$, the bond-hole boson amplitude diminishes, which
drives the charge gap to a lower bound. Upon further increasing
$g/D$, the systems transits from a superconducting  to an
insulating phase. The non-monotonic behavior occurring in the
superconducting  regime is related to the competition between the
two possible channels for breaking a $b$-bond boson (see Fig.1b).
Moving into the insulating side of the phase diagram, we notice
that the charge gap increases almost linearly with $g$. This
indicates that now, for breaking one $b$-bond boson in order to
create a pair of $s$, $h$-bond fermions, one has to overcome the
strength of the local exchange. In this limit, it is the
constraint that determines the local scale of exchange energy in
the excitation spectrum.

Up to now we have concentrated our attention on the amplitude of
the $b$-bond bosons and their, related to it, charge gap and its
evolution from the SC to the I phase. We next shall investigate
how the dispersion and the related spectral weight associated with
the $c$-type excitations of the Bogoliubov branches get modified
by tuning the exchange coupling. In Fig. \ref{SPvsg} we report the
behavior of the spectral function for the original fermions as a
function of their momentum, when moving along the major symmetry
lines of the two-dimensional Brillouin zone.

We observe that the structure of the excitation spectra is
qualitative different in the superconducting phase and the
insulating one. We obtain four branches in the SC region and only
two in the insulating one. Concerning the dispersion and the
position where the charge gap opens, there are distinct
differences in the two parts of the phase diagram. Starting from
the weak coupling regime, one  notices that the gap opens up at
the $Z$ point of the Brillouin zone. This is due to a crossing at
this point of the fermion-like and fermion-hole-like tight-binding
spectrum, for the particle-hole symmetric half filled band case,
studied here. Looking at the evolution of the spectral function,
the excitations with the dominant weight follow a dispersion that
is reminiscent of a free tight binding spectrum, except close to
the $Z$ point, where the gap opens. Furthermore, in this region of
coupling,  two almost dispersionless branches with small spectral
weights are  present. They arise from the hybridization between
the $s$- and $h$-bond fermions. Their small spectral weight is a
consequence of a counter-active action between the $b$-bond and
$d, f$-bond bosons in  renormalizing the hybridization between the
bond fermions. Increasing the exchange coupling induces a
modification in the spectral weight distribution and in the width
of the dispersion. The dominant Bogoliubov branches get more
dispersive upon approaching the quantum phase transition due to an
increase in the $b$-bond boson amplitude $\bar b$. Their spectral
weight however diminishes down to zero due to the reduction of the
bond-hole and double paired boson amplitudes $\bar d$ and $\bar f$
as we approach the quantum critical point. This change is
accompanied by a partial redistribution of the spectral weight in
the lower and upper Bogoliubov branches. Crossing the critical
exchange coupling, the bands below (above) the chemical potential
develop a unique structure and one no longer discerns any trace of
the four Bogoliubov branches which were caused by the pairing and
the hybridization between the bond-fermions. In this region, only
the process of mixing the $s$- and $h$-bond fermions is active.
Now, a $b$-bond boson can only break into a $s$- and $h$-bond
fermion but the  energy cost for that grows proportionally to the
exchange coupling and is controlled by the constraint. The
emerging bond-fermions have propagating features and hybridize
with each other which results in  a dispersion that is
renormalized by the $b$-bond boson amplitude $\bar b$.

\section{Conclusions}

In the present study we have illustrated the evolution of pairing
correlations in a two-component system where local bond pairing is
induced in an ensemble of itinerant fermions. We have shown that
there are two possible bond pair configurations which can
materialize. Depending on whether we have a finite or a zero
amplitude of the bond-hole (respectively double-paired-bond)
bosons, we obtain a superconducting or an insulating phase. The
bond fermions which emerge in such system control the occurrence
of the one or the other bond pair liquid. Those bond fermions are
both: (i) single fermions on a bond and (ii) fermions attached
locally to a boson on that same bond. The dynamics of these bond
fermions and the interplay between the related pairing and the
hybridization process crucially determines the competition between
the superconducting and the insulating paired liquid states.

To extract the main features of this scenario, we have used a bond
operator formalism, which has been widely applied in the
literature for the spin analogue quantum problems. In the present
problem, it has turned out to be particularly useful because of
the intrinsic nature of the dimer formation of resonant pairs.
Moreover, though the bond operator method, within the present mean
field approach, is best suited for the strong coupling limit, it
can indeed describe the quantum phase transition because, as we
have shown, it occurs in a regime of coupling  where $g/t \sim 2$.
This, a posteriori, justifies the use of an approximation that is
based on a local saddle point ansatz for the bond bosons. Still,
as we have discussed in the Introduction, it is already within the
local structure of the bond configurations that one recovers the
objects to distinguish between the superonducting and the
insulating state. Among the non-bonding configurations, there is
the bond-hole state, which turns out to be the relevant bond
boson, because the formation of the superconducting state is
attributed to the simultaneous presence of a phase coherent
amplitude for the bonding bosons as well as for the bond holes.
This is a key element that allows to follow the evolution from the
insulating (condensation of the pure bonding bosons) to the
superconducting state (pairing of the $s$ and $h$ fermions, due to
the phase coherence of the bonding bosons and the bond holes).

We have obtained the phase diagram as a function of the
boson-fermion exchange coupling and by varying the temperature. We
have shown how, with increasing the exchange coupling, the
continuous reduction of the amplitude of the bond-hole condensate
drives the quantum phase transition between the superconducting
and the insulating state. This behavior is dictated by the
interplay between the dynamics of the bond fermions and the
constraint controlling the occupation of the various bond
operators. In this first attempt to capture the salient aspects of
that scenario, on the basis of a mean field approach in the bond
operator formulation of it, we have focused our attention on the
excitation spectra and the nature of the charge gap. In the
superconducting region the single particle gap exhibits a
non-monotonic evolution as the exchange coupling is increased and
the quantum phase transition is approached. This is a sign of the
double nature of the dynamical variables at play and which
contains features of pairing as well as of hybridization of those
bond fermions. A completely different behavior characterizes the
insulating regime where it is the local exchange which sets the
scale of energy. Coming from the insulating side, entering the
superconducting phase one is led to the picture where the
insulating charge gap continues to exist in the superconducting
phase as one of the components of the charge gap, and whose
amplitude diminishes as $g$ goes to zero. This could signify a
breakdown in cascade of the order in such systems, as $g$ is
increased.  First destroying the superconducting phase but without
the amplitude of the pairs up to the quantum critical point and
then destroying gradually the pair amplitude when $g$ is
sufficiently large in the insulating state.

Concerning the excitation spectra, there are four Bogoliubov
branches in the superconducting region and which continuously
evolve into  a two branch configuration in the insulating  region.
For a 2-dimensional system and the corresponding tight-binding
spectrum for the original bare fermions, the pairing gap opens up
at the $Z$ point of the Brillouin zone. The interplay between the
pairing and the hybridization between the bond fermions also
controls the spectral weight redistribution between the various
Bogoliubov branches as the exchange coupling is tuned through the
quantum phase transition.

Further analysis on the issues which have been discussed here is
presently in progress and deals with such questions as the effects
expected for hole doping away from the half-filled band case and
of the feedback on the bond fermions arising from the dynamics of
the bond bosons. A major attempt in this direction will concern
the case when the system is in the proximity of the breakdown of
the coherent superconducting  state where possibly a metallic
phase of bosonic fermion pairs could exist. For that purpose the
present bond-mean-field analysis will be improved by taking into
account the phase fluctuations of the bosonic mean field coherent
states desribed here.

\end{document}